\documentclass{article}

\usepackage[utf8]{inputenc}
\usepackage{authblk}
\usepackage{hyperref}

\usepackage{amsthm}
\usepackage{aliascnt}

\usepackage{amsmath} 
\usepackage{amsfonts}
\usepackage[T1]{fontenc}
\usepackage[english]{babel}

\newtheorem{theorem}{Theorem}
\newaliascnt{definition}{theorem}
\newtheorem{definition}[definition]{Definition}
\aliascntresetthe{definition}
\newaliascnt{conjecture}{theorem}

\aliascntresetthe{conjecture}
\newaliascnt{lemma}{theorem}

\aliascntresetthe{lemma}
\newaliascnt{notation}{theorem}

\aliascntresetthe{notation}
\newaliascnt{observation}{theorem}

\aliascntresetthe{observation}
\newaliascnt{proposition}{theorem}
\newtheorem{proposition}[proposition]{Proposition}
\aliascntresetthe{proposition}
\newaliascnt{property}{theorem}

\aliascntresetthe{property}
\newaliascnt{example}{theorem}
\newtheorem{example}[example]{Example}
\aliascntresetthe{example}
\newaliascnt{remark}{theorem}
\newtheorem{remark}[remark]{Remark}
\aliascntresetthe{remark}
\newaliascnt{corollary}{theorem}
\newtheorem{corollary}[corollary]{Corollary}
\aliascntresetthe{corollary}
\newaliascnt{construction}{theorem}

\aliascntresetthe{construction}

\newtheorem*{theorem*}{Theorem}

\title{Onion De Bruijn Sequences: Fixed-Window Counting by Growing the Alphabet}
\author[]{Dor Genosar}
\author[]{Yotam Svoray}
\author[]{Gera Weiss}
\affil[]{The Stein Faculty of Computer and Information Science\\Ben-Gurion University of the Negev\\Beer-Sheva, Israel}
\affil[]{{\footnotesize\texttt{genosard@post.bgu.ac.il, ysavorai@post.bgu.ac.il, geraw@bgu.ac.il}}}
\date{ }

\begin{document}

\maketitle

\begin{abstract}
We study a fixed-window counting system in which integers are represented by words of constant length while the alphabet grows as needed. This viewpoint arises from De Bruijn sequences: for fixed order $n$, the reverse prefer-max sequence is compatible with alphabet growth, since for each $k$ its restriction to $[k]^n$ is a De Bruijn sequence, yielding an infinite sequence over $\mathbb{N}$. We formalize this through the notion of an onion De Bruijn sequence, prove the resulting structural properties, and count compatible finite onion prefixes by an explicit product formula. For orders $n=2,3$, we give explicit rank and unrank formulas and describe addition and multiplication via finite normalization, with exact carry counts and linear carry complexity in the input layers. A bounded register-level experiment further shows that moving-pointer onion implementations yield strictly local data-field updates together with improved worst-case switching and burstiness relative to binary and Gray-encoded rank counters.
\end{abstract}

\tableofcontents

\section{Introduction}
Given positive integers $n$ and $k$, a \textbf{De Bruijn sequence} of order $n$ over the alphabet $[k]=\{0, \dots, k-1\}$ is a cyclic sequence $(w_i)_{i=0}^{k^n-1}$ of words in $[k]^n$ such that 
\begin{enumerate}
    \item If $w_i = \sigma_1 \cdots \sigma_{n}$ for $ \sigma_1, \dots, \sigma_{n} \in [k]$ then $w_{i+1 \bmod k^n} = \sigma_2 \cdots \sigma_n \tau$ for some $\tau \in [k]$. 
    \item If $i \neq j$ then $w_i \neq w_j$.
\end{enumerate}

For example, for $n=2$ and $k=2$, the sequence $00, 01, 11, 10$ is a De Bruijn sequence. One can view De Bruijn sequences as Hamiltonian cycles in the De Bruijn digraph $DB(n,k)$, whose vertices are elements of $[k]^n$ and where $\sigma_1 \cdots \sigma_n$ is connected to $\tau_1 \cdots \tau_n$ if and only if $\sigma_i=\tau_{i+1}$ for every $i$. De Bruijn sequences are named after N. G. de Bruijn, who studied them systematically in~\cite{de1946combinatorial}. 

Most work on infinite or extendable De Bruijn sequences varies the order parameter $n$ while keeping the alphabet fixed. A representative result in this direction is the following theorem of Becher and Heiber~\cite{becher2011extending}:

\begin{theorem*}[Theorem $1$ in~\cite{becher2011extending}]
    Every De Bruijn sequence of order $n$ in at least three symbols can be extended to a
De Bruijn sequence of order $n + 1$ (over the same alphabet). Every De Bruijn sequence of order $n$ in two symbols
cannot be extended to order $n + 1$, but it can be extended to order $n + 2$.
\end{theorem*}

A complementary question is to keep the order $n$ fixed while increasing the alphabet size. Can a De Bruijn sequence of order $n$ over $[k]$ be extended to one over $[k+1]$ without changing the window\footnote{By "window" we mean as an element of the total De Bruijn sequence, i.e., a word of length $n$.} length? Equivalently, can one coherently enumerate the sets $[k]^n$ for all $k$ so that the $k$-th stage extends the previous one? This leads to the following definition:

\begin{definition}\label{def:onion}
We say that a sequence $(x_i)_{i=0}^\infty$ of words in $\mathbb{N}^n$ is \textbf{an onion De Bruijn sequence} of order $n$ if, for every $k\geq 1$, the prefix $(x_i)_{i=0}^{k^n-1}$ is a De Bruijn sequence of order $n$ over the alphabet $[k]$. 
\end{definition}

An onion De Bruijn sequence is therefore a coherent enumeration of the sets $[k]^n$ as the alphabet grows. The name "onion" reflects the resulting layer structure. It may be viewed as a fixed-window counting system: the register length remains $n$, and larger integers are represented by expanding the alphabet only when necessary. For this interpretation to be algorithmically useful, the successor must be computable locally from the current window, i.e.\ by a shift rule; see, for example,~\cite{amram2018bruijn2}. In the reverse prefer-max onion studied here, this is not merely a heuristic requirement: below we derive and prove such a local rule from the fixed-alphabet prefer-max shift-rule machinery.

For fixed alphabets, efficient shift-rule machinery for prefer-max and related De Bruijn sequences is already available~\cite{amram2018bruijn2,amram2019shift,amram2020gsr}; in algorithmic terms, these works provide explicit local successor logic. The new point here is not to replace that logic, but to place it inside a growing-alphabet onion.

Our motivating picture is a counter implemented by an $n$-cell shift register. The state always remains a word of length $n$: incrementing the counter means shifting the register and appending one locally determined symbol, while larger values are represented by alphabet growth rather than by widening the register. In the onion setting, that appended symbol can be supplied by the same efficient shift-rule circuitry used for the finite-alphabet prefer-max sequence, after a small layer-detection wrapper chooses the relevant alphabet size. Thus the maximal symbol immediately gives a coarse magnitude estimate through the layer, and in orders $2$ and $3$ the onion representation also supports direct addition and multiplication while remaining inside the same fixed-window state space.

Different numeration systems privilege different costs. In base $10$, for example, the length of the numeral immediately gives $\lfloor \log_{10} N\rfloor$. In an onion representation, the maximal symbol immediately determines the layer and therefore $\lfloor \sqrt[n]{N}\rfloor$. This is especially natural in settings where values are already maintained as fixed windows and updated locally, such as shift-based counters, streaming state machines, or low-switching address-generation hardware. Gray-coded instruction addressing has been proposed as a way to reduce instruction-address-bus switching in embedded processors~\cite{su1994power,mehta1996gray,guo2010shifted}, feedback-shift-register counters have been studied as program counters for high-speed FPGA processors~\cite{suggate2019cyclic}, and related De Bruijn-style constrained codes have been studied for emerging memory technologies such as racetrack memory~\cite{chee2020constrained}. The onion viewpoint is aimed at the same design space, but with an additional arithmetic structure coming from a coherent growing-alphabet De Bruijn order.

Our first main result, \autoref{the:onion}, shows that the reverse prefer-max De Bruijn sequence has the onion property, yielding a canonical infinite sequence over $\mathbb{N}$. From that theorem, together with the fixed-alphabet backward prefer-max shift rule, we derive a formal successor proposition for the infinite onion order. We then identify the structural features shared by every onion sequence, count the compatible layer orders and finite onion prefixes, and translate the layer decomposition into arithmetic consequences. For the reverse prefer-max onion sequences of orders $2$ and $3$, these consequences become explicit rank and unrank formulas together with direct addition and multiplication by finite layer carries, including exact carry counts and linear carry complexity. We also include a bounded register-level experiment showing that this locality survives a concrete encoding: on a fixed state space, moving-pointer onion representations have strictly local data-field updates and better worst-case or less bursty switching than binary and Gray-encoded rank counters, even though Gray remains optimal for average Hamming distance. This viewpoint also recasts classical Diophantine statements in onion language: in general, Fermat's equation $x^n+y^n=z^n$ becomes $0^{n-1}x \oplus_n 0^{n-1}y = 0^{n-1}z$, and in order $3$ this takes the explicit form $00x \oplus_3 00y = 00z$.

The paper follows this progression. \autoref{the:onion} proves the onion property for reverse prefer-max, and the accompanying successor proposition turns the fixed-alphabet shift-rule theory into a proved local update rule for the infinite onion sequence. \autoref{section:structure} isolates the universal layer constraints, and \autoref{section:counting} counts layer orders and compatible finite onion prefixes. The next three sections develop the arithmetic side: first in general, then explicitly in orders $2$ and $3$. The final section summarizes the picture and the main open directions.

\textbf{Notation.} Throughout, we denote $\mathbb{N}=\{0,1,2,\dots\}$. For a word $y=\sigma_1 \cdots \sigma_m \in [k]^*$, let $\mu_y=\max_i \sigma_i$. For a word $w$ of length $n$ and for $m<r \leq n$, let $w[m..r]$ denote the suffix of $w$ of length $r-m+1$.

\textbf{Acknowledgments}: We wish to thank Benjamin F. Keefer (of the Data Science program in the University of Utah) for his contributions to~\autoref{section:structure}, his enumeration of the $k$-th layer, and the enumeration of compatible onion prefixes in~\autoref{section:counting}. We wish to thank Oded Margalit for bringing to our attention the use of shift registers as program counters. 

\section{The onion theorem and onion sequences}

\subsection{The onion theorem for prefer-max}

\begin{definition} \label{def:pref-max}
	The $(k,n)$-\textbf{prefer-max De Bruijn sequence}, which we denote by $(w_i)_{i=0}^{k^n-1}$, is defined recursively as follows:
     \begin{itemize}
        \item $w_0=0^{n{-}1}{(k{-}1)}$,
        \item If $w_i = \sigma x$ for some $\sigma \in [k]$ and $x \in [k]^{n-1}$ then $w_{i+1}=x\tau$ where $\tau$ is the maximal\footnote{We view $[k]$ with the usual order, i.e. $0<1<\cdots < k-1$. } symbol in $[k]$ such that $w_j \neq x \tau$ for all $j \leq i$.
    \end{itemize}
\end{definition}

The binary prefer-max De Bruijn sequence goes back to~\cite{martin1934problem, ford1957cyclic}, and the general prefer-max De Bruijn construction has since been studied extensively; see, for example,~\cite{amram2019shift,amram2020gsr,amram2021cycle}. For fixed alphabets, efficient shift rules for prefer-max De Bruijn sequences already provide a fast successor operation~\cite{amram2019shift,amram2020gsr}. The onion setting extends that local-update viewpoint to growing alphabets and turns it into a framework for fixed-window counting and arithmetic. Our first result identifies a suffix relation between the $(k,n)$-prefer-max and the $(k-1,n)$-prefer-max sequences:

\begin{theorem}[Onion Theorem] \label{the:onion}
	For every $k>1$ and $n\geq 2$, if $(w_i)_{i=0}^{k^n-1}$ is the $(k,n)$-prefer-max sequence, then its suffix
    $(w_i)_{i=k^n-(k-1)^n}^{k^n-1}$
    is the $(k-1,n)$-prefer-max De Bruijn sequence.
\end{theorem}
\begin{proof} 
    For $n=2$, the words containing $k-1$ form the initial block
    \[
        \begin{aligned}
        &0(k-1),\; (k-1)(k-1),\; (k-1)(k-2),\; (k-2)(k-1),\; \dots,\\
        &(k-1)1,\; 1(k-1),\; (k-1)0.
        \end{aligned}
    \]
    Indeed, after $j(k-1)$ with $1\leq j\leq k-1$, the largest unseen word with prefix $k-1$ is $(k-1)(j-1)$, and after $(k-1)j$ with $1\leq j\leq k-2$, the largest unseen word with prefix $j$ is $j(k-1)$. After $(k-1)0$, the next word is $0(k-2)$, so the construction continues exactly as the $(k-1,2)$-prefer-max construction.
    
    Assume that $n>2$, and view the prefer-max cycle through a linearized symbol string $(\sigma_i)$ whose consecutive length-$n$ windows are the words $w_i$. Let
    \[
        i_0=\min\{i\colon  w_i \in [k-1]^n\}.
    \]
    We first show that
    \[
        w_{i_0-1}=(k-1)0^{n-1}
        \qquad\text{and}\qquad
        w_{i_0}=0^{n-1}(k-2).
    \]
    By minimality, the unique symbol that leaves when we pass from $w_{i_0-1}$ to $w_{i_0}$ is $k-1$, so $\sigma_{i_0-n}=k-1$. Since $\sigma_{i_0}\neq k-1$ and the word
    \[
        \sigma_{i_0-n+1}\cdots \sigma_{i_0-1}(k-2)
    \]
    does not contain $k-1$, it has not appeared earlier. Hence the prefer-max rule forces $\sigma_{i_0}=k-2$, and therefore the word
    \[
        \sigma_{i_0-n+1}\cdots \sigma_{i_0-1}(k-1)
    \]
    appears at some earlier index $i_1<i_0$.

    If $\sigma_{i_0-n+1}\cdots \sigma_{i_0-1}\neq 0^{n-1}$, then this earlier word is not $0^{n-1}(k-1)$ and therefore has a predecessor. That predecessor contains $k-1$, while its last $n-1$ symbols do not, so its first symbol must be $k-1$. Thus $w_{i_1-1}=w_{i_0-1}$, contradicting that a De Bruijn sequence has no repeated windows. Hence $\sigma_{i_0-n+1}\cdots \sigma_{i_0-1}=0^{n-1}$, proving the claim.

    We next show that every word containing $k-1$ appears before $i_0$. First suppose that $v\in [k]^n$ ends with $k-1$. Because the prefer-max rule appends $0$ only after all larger continuations of the same prefix have already appeared, the words
    \[
        v,\ v[2..n]0,\ v[3..n]0^2,\ \dots,\ v[n]0^{n-1}=(k-1)0^{n-1}
    \]
    appear in this order. Since $(k-1)0^{n-1}=w_{i_0-1}$, every word ending with $k-1$ appears before $i_0$.

    Now let $v$ be any word containing $k-1$, and let the rightmost occurrence of $k-1$ in $v$ lie in position $r$. The window that starts $r-1$ steps earlier ends with this same occurrence of $k-1$; call it $u$. By the previous paragraph, $u$ appears before $i_0$. Moreover, every window from $u$ up to $v$ still contains that fixed occurrence of $k-1$, so none of them can be $w_{i_0}$. Therefore $v$ also appears before $i_0$.

    Thus the windows before $i_0$ are exactly the words in $[k]^n\setminus [k-1]^n$. Since $w_{i_0-1}=(k-1)0^{n-1}$ and $w_{i_0}=0^{n-1}(k-2)$, the prefer-max rule from index $i_0$ onward uses only the alphabet $[k-1]$ and proceeds exactly as in the $(k-1,n)$-prefer-max construction. Hence $(w_i)_{i=i_0}^{k^n-1}$ is the $(k-1,n)$-prefer-max sequence. Since
    \[
        i_0=\left|[k]^n \setminus [k-1]^n\right|=k^n-(k-1)^n,
    \]
    this is precisely the claimed suffix.
\end{proof}

\begin{remark}
    \textup{An alternative proof of~\autoref{the:onion} is presented in Theorem 26 of~\cite{amram2021cycle} using the cycle-joining construction presented in that paper. }
\end{remark}

\begin{example}
    \textup{The $(2,2)$-prefer-max sequence is $01100$, and the $(3,2)$-prefer-max sequence is $0221201100$; the former appears as a suffix of the latter.}
\end{example}

By \autoref{the:onion}, for every $n\geq 2$ the reverse of the $(2,n)$-prefer-max sequence is a prefix of the reverse of the $(3,n)$-prefer-max sequence, which is a prefix of the reverse of the $(4,n)$-prefer-max sequence, and so on. Thus the reversed prefer-max sequences define an infinite onion De Bruijn sequence of order $n$, i.e., an infinite sequence in which every word in $\mathbb{N}^n$ appears exactly once as a contiguous block. 

This theorem also clarifies why shift-register constructions are so natural in the onion setting: compatibility across alphabet sizes becomes a local successor problem. The next proposition turns this into a precise and proved statement. It gives a shift-and-append rule for the current window and shows that the rule is inherited from the fixed-alphabet backward prefer-max shift rule once the relevant layer is identified.

\noindent
We write $>_{\mathrm{colex}}$ for right-to-left lexicographic order on words of the same length, so $a>_{\mathrm{colex}} b$ if and only if $a^{\mathrm{rev}}$ is lexicographically larger than $b^{\mathrm{rev}}$.

\begin{proposition}[Onion successor rule]\label{prop:onion_successor_rule}
    Fix $n\geq 2$. For $x\in\mathbb{N}^{n-1}$ and $\tau\in\mathbb{N}$, write $x=v0^\ell$, where $v$ is empty or ends in a nonzero symbol, and denote $\operatorname{adm}(x,\tau)$ if $\tau>0$ and $0^\ell\tau v$ is $>_{\mathrm{colex}}$-maximal among the rotations of $x\tau$. If $\sigma x$ is a state in the infinite reverse prefer-max onion sequence, then
    \[
        \operatorname{succ}(\sigma x)=
        \begin{cases}
            x(\sigma+1), & \operatorname{adm}(x,\sigma+1),\\
            x0, & \operatorname{adm}(x,\sigma)\ \text{and}\ \forall\tau>\sigma,\ \neg\operatorname{adm}(x,\tau),\\
            x\sigma, & \text{otherwise.}
        \end{cases}
    \]
\end{proposition}

\begin{proof}
    Let $k=\mu_{\sigma x}+2$. Since $\sigma x\in [k-1]^n$, \autoref{the:onion} implies that the successor of $\sigma x$ in the infinite onion sequence is its successor in the reverse of the $(k,n)$-prefer-max sequence.

    Write $x=v0^\ell$. Reversing each rotation of $x\tau$ converts colex order into ordinary lexicographic order on the rotations of $x^{\mathrm{rev}}\tau$, and under this reversal the distinguished rotation $0^\ell\tau v$ becomes $v^{\mathrm{rev}}\tau 0^\ell$. Thus $\operatorname{adm}(x,\tau)$ is exactly the predicate that appears in the backward prefer-max shift rule (as appears in Proposition~20 and Theorem~24 of~\cite{amram2018bruijn2}), rewritten in our notation.

    For fixed $x$, the admissible positive symbols form an initial interval. Indeed, if $\operatorname{adm}(x,\tau)$ holds and $0<\tau'<\tau$, let $d_\tau=0^\ell\tau v$ and let $r_\tau$ be any other rotation of $x\tau$. In the colex comparison between $d_\tau$ and $r_\tau$, the first differing position either is independent of $\tau$, in which case the inequality is unchanged when $\tau$ is replaced by $\tau'$, or else it contains $\tau$ in $d_\tau$ and a smaller symbol in $r_\tau$, in which case replacing $\tau$ by $\tau'$ still leaves $d_{\tau'}$ larger than $r_{\tau'}$. Hence $\operatorname{adm}(x,\tau')$ also holds.

    The fixed-alphabet backward shift rule from~\cite[Theorem~24]{amram2018bruijn2} therefore yields three cases for the successor of $\sigma x$ in the reverse of the $(k,n)$-prefer-max sequence: append $\sigma+1$ when $\operatorname{adm}(x,\sigma+1)$ holds; append $0$ when $\sigma$ is the largest admissible symbol; otherwise append $\sigma$. Since the admissible symbols form an initial interval, the middle condition is exactly
    $
        \operatorname{adm}(x,\sigma)\ \text{and}\ \forall\tau>\sigma,\ \neg\operatorname{adm}(x,\tau).
    $
    This is the displayed rule.
\end{proof}

\noindent
Thus, once the current layer determines the ambient alphabet $[\,\mu_{\sigma x}+2\,]$, the onion successor is exactly the fixed-alphabet backward prefer-max shift rule rewritten in colex form.

\begin{example}[Counting with the onion successor]
    \textup{We record three states of length $7$, one for each branch of \autoref{prop:onion_successor_rule}.}

    \textup{First take $\sigma x=0001316$. Then $\sigma=0$, $x=001316$, and $\mu_{\sigma x}=6$. The distinguished rotations $\tau001316$ are colex-maximal for $\tau=1,\dots,5$, so $\operatorname{adm}(001316,\tau)$ holds for those values. By contrast, $\operatorname{adm}(001316,6)$ fails because $0013166>_{\mathrm{colex}}6001316$, and therefore no larger symbol is admissible either. Since $\sigma+1=1$ is admissible, the first branch gives}
    \[
        \operatorname{succ}(0001316)=0013161.
    \]

    \textup{Next let $\sigma x=6413067$. Then $\sigma=6$, $x=413067$, and $\mu_{\sigma x}=7$. Here $\operatorname{adm}(413067,\tau)$ holds for $\tau=1,\dots,6$; indeed, the rotations $\tau413067$ are colex-maximal in their rotation classes. But $\operatorname{adm}(413067,7)$ fails, since the rotation $4130677$ is colex larger than the distinguished rotation $7413067$. Hence $\sigma=6$ is the largest admissible symbol, and the second branch yields}
    \[
        \operatorname{succ}(6413067)=4130670.
    \]

    \textup{Finally consider $\sigma x=2100000$. Then $\sigma=2$, $x=100000$, and $\mu_{\sigma x}=2$, so the positive candidates are $\tau\in\{1,2,3\}$. For $\tau=1$, the distinguished rotation is $0000011$, and it is colex-maximal among the rotations of $1000001$, so $\operatorname{adm}(100000,1)$ holds. For $\tau=2$, the distinguished rotation would be $0000021$, but the rotation $1000002$ is larger in colex order, so $\operatorname{adm}(100000,2)$ fails; hence no larger symbol is admissible. Thus there is an admissible positive symbol, but it lies below $\sigma$. Neither of the first two cases applies, and the third branch gives}
    \[
        \operatorname{succ}(2100000)=1000002.
    \]

    \textup{The last example shows that the branch $\operatorname{succ}(\sigma x)=x\sigma$ does not require the admissible set to be empty. It applies whenever the largest admissible positive symbol is strictly smaller than $\sigma$.}
\end{example}

\subsection{Structure of onion De Bruijn sequences}\label{section:structure}

By \autoref{the:onion}, the reverse prefer-max construction gives one onion De Bruijn sequence. This section records the structural features shared by all of them. First we show that, once a context of length $n-1$ is fixed, enlarging a symbol beyond the current maximum preserves the order. Then we identify the resulting layer decomposition by maximal symbol.

\begin{proposition}[Monotonicity in a fixed context]\label{prop:single_digit_monotone}
    Let $(x_i)_{i=0}^\infty$ be an onion De Bruijn sequence of order $n$, and let $w$ and $u$ be two words whose lengths sum to $n-1$. Then the subsequence $\{w\sigma u \colon \sigma > \mu_{wu} \}$ appears in increasing order in $(x_i)_{i=0}^\infty$.
\end{proposition}

\begin{proof}
    Since $(x_i)_{i=0}^\infty$ is an onion De Bruijn sequence, the set $\{w\sigma u \colon \sigma\leq \mu_{wu}\}$ already appears in the $(\mu_{wu}+1,n)$-De Bruijn prefix of $(x_i)_{i=0}^\infty$. Assume towards contradiction that there exist symbols $\mu_{wu}<\sigma_1<\sigma_2$ such that $x_{i_1}=w\sigma_2u$ and $x_{i_2}=w\sigma_1u$ with $i_1<i_2$. Since $w\sigma_1u\in [\sigma_1+1]^n$, the word $w\sigma_1u$ appears in the prefix $(x_i)_{i=0}^{(\sigma_1+1)^n-1}$. On the other hand, $w\sigma_2u\notin [\sigma_1+1]^n$, so its unique occurrence cannot lie in that prefix. Hence $i_2\leq (\sigma_1+1)^n-1<i_1$, a contradiction.
\end{proof}

This monotonicity reflects a more global rigidity. In every onion De Bruijn sequence, the words whose maximal symbol is $k-1$ form a single contiguous block, namely the interval added when one passes from the $(k-1,n)$ prefix to the $(k,n)$ prefix. The next proposition makes this layer decomposition precise.

\begin{proposition}\label{prop:layer_decomposition}
    Let $(x_i)_{i=0}^\infty$ be an onion De Bruijn sequence of order $n$. Then for every $k \geq 2$ and every index $i\geq 0$,
    \[
        x_i \in [k]^n \setminus [k-1]^n
        \quad\Longleftrightarrow\quad
        (k-1)^n \leq i \leq k^n-1.
    \]
\end{proposition}

\begin{proof}
    Let $k\geq 2$. First, assume that $(k-1)^n \leq i \leq k^n-1$. Since $(x_i)_{i=0}^\infty$ is an onion De Bruijn sequence of order $n$, the prefix $(x_i)_{i=0}^{(k-1)^n-1}$ is an $(n,k-1)$-De Bruijn sequence and the prefix $(x_i)_{i=0}^{k^n-1}$ is an $(n,k)$-De Bruijn sequence. Thus every word in $[k-1]^n$ already appears exactly once among $x_0,\dots,x_{(k-1)^n-1}$. Since the longer prefix is a De Bruijn sequence over $[k]$, we also have $x_i \in [k]^n$. Because $i \geq (k-1)^n$, the word $x_i$ cannot belong to $[k-1]^n$, and therefore $x_i \in [k]^n \setminus [k-1]^n$.

    Conversely, assume that $x_i \in [k]^n \setminus [k-1]^n$. The prefix
    $(x_i)_{i=0}^{k^n-1}$ is an $(n,k)$-De Bruijn sequence, so every word in
    $[k]^n$ appears exactly once among $x_0,\dots,x_{k^n-1}$. Hence
    $i\leq k^n-1$. On the other hand, the prefix
    $(x_i)_{i=0}^{(k-1)^n-1}$ already contains every word in $[k-1]^n$
    exactly once, and $x_i \notin [k-1]^n$. Therefore
    $i\geq (k-1)^n$, and so $(k-1)^n \leq i \leq k^n-1$.
\end{proof}

\subsection{Counting layer orders and onion prefixes}\label{section:counting}

The layer decomposition also has an enumerative side. For fixed $n$ and $k$, let $L_{n,k}$ denote the induced subgraph of the De Bruijn digraph $DB(n,k)$ on the vertex set $[k]^n \setminus [k-1]^n$. A Hamiltonian cycle in $L_{n,k}$ is exactly an ordering of the $k$-th layer that respects the De Bruijn overlap condition.

\begin{theorem}[Enumeration of the $k$-th layer]\label{thm:layer_enumeration}
    For every $n\geq 1$ and $k\geq 1$, the number of Hamiltonian cycles in $L_{n,k}$ is
    \[
        \frac{k!^{\,k^{n-1}-(k-1)^{n-1}}}{k^{n-1}}.
    \]
\end{theorem}

\begin{proof}
    The case $n=1$ is immediate, since $L_{1,k}$ consists of the single vertex $k-1$. Assume therefore that $n\geq 2$, and write $m=n-1$. Consider the digraph $G_{n,k}$ whose vertex set is $[k]^m$, and in which $x\tau \longrightarrow \sigma x$ is an edge whenever $\sigma x\tau \in [k]^n\setminus [k-1]^n$. The edges of $G_{n,k}$ are naturally indexed by the vertices of $L_{n,k}$ and two edges $x\tau \to \sigma x$ and $y\eta \to \varphi y$ are consecutive if and only if $\sigma x=y\eta$, which is exactly the condition that the corresponding layer words $\sigma x\tau$ and $\varphi y\eta$ overlap in $n-1$ symbols. Hence the line digraph of $G_{n,k}$ is precisely $L_{n,k}$ and therefore Hamiltonian cycles in $L_{n,k}$ are in bijection with Eulerian circuits in $G_{n,k}$.

    Note that the digraph $G_{n,k}$ is Eulerian since if a vertex $u\in [k]^m$ contains the symbol $k-1$, then any symbol of $[k]$ may be prepended or appended while staying inside $[k]^n\setminus [k-1]^n$, so $\deg_{\textup{in}}(u)=\deg_{\text{out}}(u)=k$, and if $u\in [k-1]^m$, then only the symbol $k-1$ may be prepended or appended, and therefore $\deg_{\textup{in}}(u)=\deg_{\text{out}}(u)=1$. In addition, the digraph is also strongly connected, i.e. from any vertex one reaches $(k-1)^m$ by repeatedly prepending $k-1$, and from $(k-1)^m$ one reaches any prescribed vertex by prepending its symbols from right to left.

    Denote $S=[k-1]^m$ and $L=[k]^m\setminus [k-1]^m$ and let $\mathcal{L}$ be the out-degree Laplacian of $G_{n,k}$. With respect to the decomposition $[k]^m=S\sqcup L$, the Laplacian has the block form
    \[
        \mathcal{L}=
        \begin{pmatrix}
            I_{|S|} & -A_{SL}\\
            -A_{LS} & kI_{|L|}-A_{LL}
        \end{pmatrix},
    \]
    where $A_{SL},A_{LS},A_{LL}$ are the corresponding adjacency blocks. Taking the Schur complement of the $S$-block in $\lambda I-\mathcal{L}$ gives us (after simplification) that the characteristic polynomial of $\mathcal{L}$ is
    \[
        \chi_{\mathcal{L}}(\lambda)=\lambda(\lambda-1)^{|S|-1}(\lambda-k)^{|L|},
    \]
    \noindent and therefore the nonzero Laplacian eigenvalues are $1$ with multiplicity $|S|-1$ and $k$ with multiplicity $|L|$. Now, fixing the vertex $r=(k-1)^m$, since $G_{n,k}$ is strongly connected and Eulerian, the directed Matrix-Tree theorem (see, for example,~\cite{tutte1948dissection,margoliash2010matrix}) tells us that the number of arborescences in $G_{n,k}$ rooted at $r$ is given by  
    \[
        t_r(G_{n,k})
        =\frac{1}{|[k]^m|}\prod_{\mu\neq 0}\mu
        =\frac{1}{k^m}\,1^{|S|-1}k^{|L|}
        =k^{|L|-m},
    \]
    \noindent where the product runs over all nonzero eigenvalues of $\mathcal{L}$. Since $|L|=k^m-(k-1)^m$ we can conclude that $t_r(G_{n,k})=k^{k^m-(k-1)^m-m}$.  Finally, by the BEST theorem~\cite{tutte1941unicursal,van1951circuits}, the number of Eulerian circuits in $G_{n,k}$ is
    \[
        t_r(G_{n,k})\prod_{u\in [k]^m}(\deg_{\text{out}}(u)-1)!.
    \]
    Since $(\deg_{\text{out}}(u)-1)!= 0!=1$ for every $u \in S$ and $(\deg_{\text{out}}(u)-1)!=(k-1)!$ for every $u \in L$, we get that the number of Hamiltonian cycles in $L_{n,k}$ is exactly 
    \[
        k^{k^m-(k-1)^m-m}(k-1)!^{\,k^m-(k-1)^m}=\frac{k!^{\,k^{n-1}-(k-1)^{n-1}}}{k^{n-1}},
    \]
    \noindent as desired. 
\end{proof}

\begin{example}
    \textup{In the case $n=2$, \autoref{thm:layer_enumeration} tells us that the number of Hamiltonian cycles in $L_{2,k}$ is $(k-1)!$, and we can see this directly as a Hamiltonian cycle in $L_{n,k}$ (without loss of generality) must be of the form}
    \[
        0(k-1) \to (k-1)j_1 \to  j_1(k-1) \to (k-1)j_2 \to \cdots \to j_{k-1}(k-1) \to (k-1)0,
    \]
    \noindent \textup{where $j_1, \dots, j_{k-1}$ is any re-ordering of $1, \dots, (k-1)$. In particular, the reverse of the prefer-max sequence is the case where $j_i=i$, as we see in~\autoref{the:onion}. }
\end{example}

\begin{corollary}[Enumeration of compatible onion prefixes]\label{cor:onion_prefix_count}
    For every $n\geq 1$ and $j\geq 1$, the number of finite sequences $(x_i)_{i=0}^{j^n-1}$ such that, for each $1\leq k\leq j$, the prefix $(x_i)_{i=0}^{k^n-1}$ is a De Bruijn sequence of order $n$ over the alphabet $[k]$ is
    \[
        \prod_{k=2}^{j}\frac{k!^{\,k^{n-1}-(k-1)^{n-1}}}{k^{n-1}},
    \]
    where the empty product is interpreted as $1$.
\end{corollary}

\begin{proof}
    The case $j=1$ is trivial, since the only such sequence is $(0^n)$. Assume $j\geq 2$.

    Fix $2\leq k\leq j$. Since the prefix of length $(k-1)^n$ already contains every word of $[k-1]^n$ exactly once, while the prefix of length $k^n$ contains every word of $[k]^n$ exactly once, the block
    \[
        x_{(k-1)^n},x_{(k-1)^n+1},\dots,x_{k^n-1}
    \]
    consists exactly of the words in $[k]^n\setminus [k-1]^n$.

    The last word of the length-$k^n$ prefix must overlap the first word $x_0=0^n$ in the cyclic De Bruijn order, so it has the form $\tau 0^{n-1}$. Because it lies in $[k]^n\setminus [k-1]^n$, we must have that $\tau=k-1$, and so $x_{k^n-1}=(k-1)0^{n-1}$. Similarly, the first word of the $k$-th layer must overlap the last word of the $(k-1)$-st prefix, which is $(k-2)0^{n-1}$, and must lie in $[k]^n\setminus [k-1]^n$, and therefore $x_{(k-1)^n}=0^{n-1}(k-1)$. Consequently the $k$-th layer is a Hamiltonian path in $L_{n,k}$ from $0^{n-1}(k-1)$ to $(k-1)0^{n-1}$.

    These two endpoints are adjacent in $L_{n,k}$, because both belong to $[k]^n\setminus [k-1]^n$ and $(k-1)0^{n-1}\to 0^{n-1}(k-1)$ is a valid De Bruijn overlap. Hence Hamiltonian paths in $L_{n,k}$ from $0^{n-1}(k-1)$ to $(k-1)0^{n-1}$ are in bijection with Hamiltonian cycles in $L_{n,k}$. By \autoref{thm:layer_enumeration}, there are exactly $\frac{k!^{\,k^{n-1}-(k-1)^{n-1}}}{k^{n-1}}$ choices for the $k$-th layer.

    The choices for different values of $k$ are independent, because the boundary words between consecutive layers are forced. Concatenating the chosen layers therefore produces exactly one compatible prefix of length $j^n$, and every such prefix arises uniquely in this way. Multiplying over $k=2,\dots,j$ gives the formula.
\end{proof}

\autoref{thm:layer_enumeration} and \autoref{cor:onion_prefix_count} show that onion sequences are far from unique: each layer admits many Hamiltonian orders, and these choices are independent once the boundary words are fixed. Reverse prefer-max is therefore distinguished not by uniqueness, but by the extra arithmetic regularity of one particular compatible choice.

\section{Arithmetic properties of onion representations}

Let $(x_i)_{i=0}^\infty$ be an onion De Bruijn sequence of order $n$. Since every word in $\mathbb{N}^n$ appears exactly once, we may define its \textbf{rank} by $\rho(w)=i$ if and only if $x_i=w$. This turns the onion sequence into a representation of the nonnegative integers by words of fixed length $n$.

The first observation is that the maximal symbol determines the represented magnitude exactly: the $k$-th layer occupies the interval between two consecutive perfect $n$th powers.

\begin{proposition}\label{prop:rank_layer}
    For every word $w \in \mathbb{N}^n$ we have
    $
        \mu_w^n \leq \rho(w) < (\mu_w+1)^n.
    $
    Equivalently,    
    $
        \mu_w=\lfloor \rho(w)^{1/n}\rfloor.
    $
\end{proposition}

\begin{proof}
    Let $k=\mu_w+1$. Then $w \in [k]^n \setminus [k-1]^n$. By \autoref{prop:layer_decomposition}, the words in $[k]^n \setminus [k-1]^n$ are precisely the words appearing in the interval of indices $[(k-1)^n,k^n-1]$. Therefore
    $
        (k-1)^n \leq \rho(w) \leq k^n-1.
    $
    Substituting $k=\mu_w+1$ gives the claim.
\end{proof}

Thus the maximal symbol plays the role of a most-significant digit, except that the thresholds are the perfect $n$th powers $0^n,1^n,2^n,\dots$ because the word length is fixed and the alphabet is what grows.

\begin{corollary}[Root extraction from the layer]\label{cor:layer_root}
    For every word $w \in \mathbb{N}^n$,
    \[
        \Bigl\lfloor \sqrt[n]{\rho(w)}\Bigr\rfloor=\mu_w.
    \]
    Thus the integer part of the $n$th root of the represented number can be read directly from the maximal symbol of the word.
\end{corollary}

\begin{proof}
    This is an immediate reformulation of \autoref{prop:rank_layer}.
\end{proof}

\begin{corollary}[Layer carries]\label{cor:layer_carries}
    Let $w \in \mathbb{N}^n$ and let $m \in \mathbb{N}$. If $z=\rho^{-1}(\rho(w)+m)$, then $\mu_z$ is the unique integer $t$ satisfying
    \[
        t^n \leq \rho(w)+m < (t+1)^n.
    \]
    In particular, under the successor map the maximal symbol changes precisely when one crosses a perfect $n$th-power boundary.
\end{corollary}

\begin{proof}
    This is an immediate reformulation of \autoref{prop:rank_layer}.
\end{proof}

\begin{corollary}[Dominant-digit monotonicity]\label{cor:dominant_digit}
    Let $w$ and $u$ be two words whose lengths sum to $n-1$. Then for every $\sigma>\mu_{wu}$ we have
    \[
        \sigma^n \leq \rho(w\sigma u) < (\sigma+1)^n,
    \]
    and the map $\sigma \mapsto \rho(w\sigma u)$ is strictly increasing on $\{\sigma \in \mathbb{N} \colon \sigma>\mu_{wu}\}$.
\end{corollary}

\begin{proof}
    Since $\sigma>\mu_{wu}$, we have $\mu_{w\sigma u}=\sigma$, so the interval bound follows from \autoref{prop:rank_layer}. The monotonicity statement follows from \autoref{prop:single_digit_monotone}.
\end{proof}

Together, \autoref{prop:rank_layer}, \autoref{cor:layer_root}, \autoref{cor:layer_carries}, and \autoref{cor:dominant_digit} isolate three arithmetic features common to every onion De Bruijn sequence: the layer gives immediate root extraction, carries between layers occur at predictable indices, and once one symbol dominates the rest of the word, increasing it moves forward in the represented order. The next two examples illustrate these points before we specialize to the reverse prefer-max sequence of order $2$.

\begin{example}\label{ex:order2_layers}
    \textup{To see the layer boundary concretely, consider the onion De Bruijn sequence of order $2$ obtained from the reverse prefer-max construction. Its first terms are}
    \[
        00,01,11,10,02,21,12,22,20,03,31,13,32,23,33,30,\dots
    \]
    \textup{Thus}
    \[
        \rho(12)=6,\qquad \rho(20)=8,\qquad \rho(03)=9,\qquad \rho(31)=10.
    \]
    \textup{In particular, $\mu_{12}=2$ and $2^2 \leq 6 < 3^2$, while $\mu_{31}=3$ and $3^2 \leq 10 < 4^2$, illustrating \autoref{prop:rank_layer}. The transition from $20$ to $03$ crosses the boundary $3^2=9$, so this is exactly a layer carry in the sense of \autoref{cor:layer_carries}.}
\end{example}

\begin{example}\label{ex:order2_monotone}
    \textup{To isolate dominant-digit monotonicity, keep the surrounding context fixed and vary only the symbol that exceeds it. In the same sequence,}
    \[
        \rho(10)=3<\rho(20)=8<\rho(30)=15<\rho(40)=24
    \]
    \textup{and also}
    \[
        \rho(12)=6<\rho(13)=11<\rho(14)=18.
    \]
    \textup{The first chain corresponds to the case where $w=\varepsilon$ and $u=0$, while the second corresponds to the case $w=1$ and $u=\varepsilon$.}
\end{example}

\subsection{Direct arithmetic in order \texorpdfstring{$2$}{2}}

The order-$2$ case already exposes a nontrivial carry mechanism. Inside layer $k$, the new symbol $k$ alternates between the right and left positions until the layer is exhausted, and only then does the sequence move to layer $k+1$. Thus, for order $2$, one can describe not only \emph{when} a layer carry occurs, but also the entire intra-layer motion leading to it. This is the first place where the abstract onion structure becomes an explicit dynamical rule. The next proposition makes this precise by identifying the complete order of the $k$-th layer and the terminal transition that produces the carry to the next layer.

\begin{proposition}[Explicit layer order for the reverse prefer-max sequence of order $2$]\label{prop:order2_explicit}
    Let $(x_i)_{i=0}^{\infty}$ be the onion De Bruijn sequence of order $2$ obtained from the reverse prefer-max construction, and let $\rho$ be its rank map. Then for every $k \geq 1$, the words of layer $k$ appear in the order
    \[
        0k,\; k1,\; 1k,\; k2,\; 2k,\; \dots,\; k(k-1),\; (k-1)k,\; kk,\; k0.
    \]
    Equivalently,
    \[
        \rho(jk)=k^2+2j \qquad (0\leq j < k)
    \]
    and
    \[
        \rho(kj)=k^2+2j-1 \qquad (1\leq j \leq k).
    \]
    Moreover,
    \[
        \rho(k0)=k^2+2k=(k+1)^2-1.
    \]
\end{proposition}

\begin{proof}
    Consider the forward prefer-max De Bruijn sequence of order $2$ over the alphabet $[k+1]$. We claim that its initial block of words containing the symbol $k$ is
    \[
        0k,\; kk,\; k(k-1),\; (k-1)k,\; k(k-2),\; (k-2)k,\; \dots,\; k1,\; 1k,\; k0.
    \]
    Indeed, the first word is $0k$ by definition. Since the current suffix is then $k$, the prefer-max rule chooses $kk$ next. More generally, after the word $jk$ with $1\leq j\leq k$, the next word must have prefix $k$, and among the candidates with that prefix the largest unseen one is $k(j-1)$. Likewise, after the word $kj$ with $1\leq j\leq k-1$, the next word must have prefix $j$, and the largest unseen choice is $jk$. This determines the whole block.

    By the Onion Theorem, the reverse prefer-max onion sequence is obtained by reversing these layers. For words of length $2$, reversing a word simply exchanges its two symbols, so the corresponding layer in the reverse prefer-max onion sequence is
    \[
        0k,\; k1,\; 1k,\; k2,\; 2k,\; \dots,\; k(k-1),\; (k-1)k,\; kk,\; k0.
    \]
    Since layer $k$ begins at index $k^2$, the displayed rank formulas follows.
\end{proof}

\autoref{prop:order2_explicit} is relevant because it makes layer carries completely explicit in order $2$. It identifies not only the boundary between one layer and the next, but also the exact path followed inside each layer. Indeed, inside layer $k$ the successor map alternates between the two positions:
\[
    jk \mapsto k(j+1) \quad (0\leq j<k), \qquad kj \mapsto jk \quad (1\leq j<k),
\]
followed by the two terminal transitions
\[
    kk \mapsto k0, \qquad k0 \mapsto 0(k+1).
\]
Thus the actual carry to the next alphabet layer happens only at the final step $k0 \mapsto 0(k+1)$.

\begin{example}\label{ex:order2_explicit}
    \textup{The new issue addressed by \autoref{prop:order2_explicit} is not where a layer begins, but how the successor moves \emph{inside} that layer. For example, when $k=5$ the rule gives}
    \[
        05 \to 51 \to 15 \to 52 \to 25 \to 53 \to 35 \to 54 \to 45 \to 55 \to 50 \to 06.
    \]
    \textup{Thus the words alternate between ending in $5$ and beginning with $5$ until the terminal pair $55,50$. Every step before $50\to 06$ stays inside layer $5$, and the only actual carry to the next layer is the final transition $50\to 06$.}
\end{example}

The previous proposition gives more than a description of the successor map: it yields an explicit arithmetic for the order-$2$ reverse prefer-max onion sequence. Once the rank map and its inverse are known in closed form, addition, multiplication, and Euclidean division can be transported from the ordinary arithmetic of integers to the onion representation. This gives a complete and exact arithmetic on the represented words, though not a local digit-by-digit algorithm of the usual decimal type. The resulting formulas are easy to parse conceptually: the square $m^2$ identifies the layer, while the offset from $m^2$ tells us where we are inside that layer.

\begin{corollary}[Explicit rank and unrank for order $2$]\label{cor:order2_rank_unrank}
    Let $\rho_2$ be the rank map of the reverse prefer-max onion sequence of order $2$. Then for every $a,b \in \mathbb{N}$ we have
    \[
        \rho_2(a,b)=
        \begin{cases}
            b^2+2a, & a<b,\\
            a^2+2b-1, & 0<b\leq a,\\
            a^2+2a, & b=0.
        \end{cases}
    \]
    Conversely, if $N \in \mathbb{N}$, $m=\lfloor \sqrt{N}\rfloor$, and $t=N-m^2$, then
    \[
        \rho_2^{-1}(N)=
        \begin{cases}
            \left(\frac{t}{2},m\right), & t \text{ is even and } t<2m,\\
            \left(m,\frac{t+1}{2}\right), & t \text{ is odd},\\
            (m,0), & t=2m.
        \end{cases}
    \]
    Consequently, the operations
    \[
        (a,b)\oplus_2(c,d)=\rho_2^{-1}\bigl(\rho_2(a,b)+\rho_2(c,d)\bigr)
    \]
    and
    \[
        (a,b)\otimes_2(c,d)=\rho_2^{-1}\bigl(\rho_2(a,b)\rho_2(c,d)\bigr)
    \]
    are explicit addition and multiplication rules on the order-$2$ onion representations. Moreover, for $(c,d)\neq (0,0)$,
    \[
        Q_2((a,b),(c,d))=\rho_2^{-1}\left(\left\lfloor \frac{\rho_2(a,b)}{\rho_2(c,d)}\right\rfloor\right)
    \]
    and
    \[
        R_2((a,b),(c,d))=\rho_2^{-1}\bigl(\rho_2(a,b)\bmod \rho_2(c,d)\bigr)
    \]
    are explicit quotient and remainder rules on the order-$2$ onion representations.
\end{corollary}

\begin{proof}
    The formula for $\rho_2$ is a restatement of \autoref{prop:order2_explicit}. The inverse formula follows by writing $N=m^2+t$ with $0\leq t\leq 2m$ and reading off the unique word in layer $m$ having offset $t$ in the list from \autoref{prop:order2_explicit}. The formulas for $\oplus_2$, $\otimes_2$, $Q_2$, and $R_2$ are then immediate.
\end{proof}

\autoref{cor:order2_rank_unrank} transports ordinary addition, multiplication, quotient, and remainder to the order-$2$ onion representation through the rank map. We now rewrite addition and multiplication directly in layer-offset coordinates.

By \autoref{prop:order2_explicit}, the $m$-th layer is a zig-zag path
\[
    0m,\; m1,\; 1m,\; m2,\; 2m,\; \dots,\; m(m-1),\; (m-1)m,\; mm,\; m0.
\]
Accordingly, write
\[
    \lambda_2(a,b)=
    \begin{cases}
        (b,a,0), & a<b,\\
        (a,a,0), & b=0,\\
        (a,b,1), & 0<b\leq a.
    \end{cases}
\]
If $\lambda_2(x)=(m,u,\varepsilon)$, then $\rho_2(x)=m^2+2u-\varepsilon$, where $\varepsilon\le u\le m$ and $\varepsilon\in\{0,1\}$. Thus $m=\lfloor \sqrt{\rho_2(x)}\rfloor$ records the layer, $u$ the offset inside the layer, and $\varepsilon$ which branch of the zig-zag one is on.

\begin{proposition}[Direct addition for order $2$]\label{prop:order2_addition_nosqrt}
    Let $\lambda_2(x)=(m,u,\varepsilon)$ and $\lambda_2(y)=(n,v,\eta)$. Define the raw sum triple by
    \[
        M=m+n,\qquad U=u+v-mn-\Bigl\lfloor \frac{\varepsilon+\eta}{2}\Bigr\rfloor,\qquad E=(\varepsilon+\eta)\bmod 2.
    \]  
    Then
    \[
        \rho_2(x)+\rho_2(y)=M^2+2U-E.
    \]
    In the special case $\lambda_2(x)=(b,a,0)$ and $\lambda_2(y)=(d,c,0)$, this reduces to the identity $M=b+d$, $U=a+c-bd$, $E=0$.
    Moreover $U\leq M$, so the raw triple can fail to be canonical only on the low side. In that case one repeatedly applies the downward carry rule
    \[
        (m,u,\varepsilon)\longmapsto (m-1,u+m-\varepsilon,1-\varepsilon)\qquad \text{if }u+1-\varepsilon\leq0,
    \]
    This process terminates at a unique canonical triple $(m,u,\varepsilon)$, and the corresponding sum word is
    \[
        x\oplus_2 y=
        \begin{cases}
            (u,m), & \varepsilon=0 \text{ and } u<m,\\
            (m,0), & \varepsilon=0 \text{ and } u=m,\\
            (m,u), & \varepsilon=1.
        \end{cases}
    \]
    Hence order-$2$ addition can be carried out directly in the onion representation, without global rank inversion.
\end{proposition}

\begin{proof}
    Adding the identities $\rho_2(x)=m^2+2u-\varepsilon$ and $\rho_2(y)=n^2+2v-\eta$ gives
    \[
        \rho_2(x)+\rho_2(y)=m^2+n^2+2u+2v-\varepsilon-\eta
        =(m+n)^2+2\Bigl(u+v-mn-\Bigl\lfloor \frac{\varepsilon+\eta}{2}\Bigr\rfloor\Bigr)-E,
    \]
    which is the claimed raw-sum formula. Since $u\leq m$ and $v\leq n$, we have
    \[
        U\leq m+n-mn\leq m+n=M,
    \]
    so the raw triple cannot overshoot the top of layer $M$.

    The two carry rules are the identities
    \[
        m^2+2u=(m-1)^2+2(u+m)-1
    \]
    and
    \[
        m^2+2u-1=(m-1)^2+2(u+m-1),
    \]
    so each carry preserves the represented integer. If $u<0$ in the even case then $m^2+2u<m^2$, while if $u\leq 0$ in the odd case then $m^2+2u-1<m^2$. In either situation, the represented integer lies strictly below layer $m$, so the canonical representative must occur in a smaller layer. Each carry lowers the layer by $1$, hence the process must terminate. The final triple is unique because $\rho_2^{-1}$ is unique.
\end{proof}

\begin{proposition}[Direct multiplication for order $2$]\label{prop:order2_multiplication_nosqrt}
    Let $\lambda_2(x)=(m,u,\varepsilon)$ and $\lambda_2(y)=(n,v,\eta)$. Define the raw product triple by
    \[
        M=mn,\qquad S=\eta m^2+\varepsilon n^2-\varepsilon\eta,
    \]
    \[
        U=m^2v+n^2u+2uv-u\eta-v\varepsilon-\Bigl\lfloor\frac{S}{2}\Bigr\rfloor,\qquad E=S\bmod 2.
    \]
    Then
    \[
        \rho_2(x)\rho_2(y)=M^2+2U-E.
    \]
    Since $\rho_2(x)\geq m^2$ and $\rho_2(y)\geq n^2$, the raw triple cannot lie below layer $M$. Thus it can fail to be canonical only on the high side, in which case one repeatedly applies the upward carry rule
    \[
        (m,u,\varepsilon)\longmapsto (m+1,u-m-\varepsilon,1-\varepsilon)\qquad \text{while }u>m
    \]
    This process terminates at a unique canonical triple $(m,u,\varepsilon)$, and the corresponding product word is
    \[
        x\otimes_2 y=
        \begin{cases}
            (u,m), & \varepsilon=0 \text{ and } u<m,\\
            (m,0), & \varepsilon=0 \text{ and } u=m,\\
            (m,u), & \varepsilon=1.
        \end{cases}
    \]
    Hence order-$2$ multiplication can also be carried out directly in the onion representation, without global rank inversion.
\end{proposition}

\begin{proof}
    Writing $\rho_2(x)=m^2+2u-\varepsilon$ and $\rho_2(y)=n^2+2v-\eta$, we obtain
    \[
        \rho_2(x)\rho_2(y)=(mn)^2+2(m^2v+n^2u+2uv-u\eta-v\varepsilon)-S.
    \]
    Since $S=2\lfloor S/2\rfloor +(S\bmod 2)$, this becomes
    \[
        \rho_2(x)\rho_2(y)=M^2+2U-E,
    \]
    which is the desired raw product formula. Moreover,
    \[
        \rho_2(x)\rho_2(y)\geq m^2n^2=M^2,
    \]
    so the represented integer cannot lie below layer $M$.

    The two upward carry rules are the identities
    \[
        m^2+2u=(m+1)^2+2(u-m)-1
    \]
    and
    \[
        m^2+2u-1=(m+1)^2+2(u-m-1),
    \]
    so each carry preserves the represented integer. If $u>m$ in the even case then $m^2+2u>(m+1)^2-1$, while if $u>m$ in the odd case then $m^2+2u-1\geq (m+1)^2$. Hence the canonical representative lies in a larger layer whenever a carry is applied. After a carry, the excess over the top of the layer strictly decreases:
    \[
        (u-m)\longmapsto u-2m-1
    \]
    in the even case and
    \[
        (u-m)\longmapsto u-2m-2
    \]
    in the odd case. Therefore the process terminates. The final triple is unique because $\rho_2^{-1}$ is unique.
\end{proof}

\autoref{cor:order2_rank_unrank} gives a closed rank and unrank description, while \autoref{prop:order2_addition_nosqrt} and \autoref{prop:order2_multiplication_nosqrt} show that addition and multiplication can be performed directly in the representation by finite layer carries. In the reverse prefer-max order, the layers are therefore not just a coarse size classification: they also provide the normalization mechanism that turns raw algebraic data back into canonical words. The next example illustrates both carry-based viewpoints.

\begin{example}\label{ex:order2_add_mult}
    \textup{In the order-$2$ reverse prefer-max onion sequence we have}
    \[
        \rho_2(24)=20,\qquad \rho_2(31)=10.
    \]
    For addition, \autoref{prop:order2_addition_nosqrt} gives
    \[
        \lambda_2(24)=(4,2,0),\qquad \lambda_2(31)=(3,1,1).
    \]
    Therefore the raw sum triple is
    \[
        (M,U,E)=(7,2+1-4\cdot 3,1)=(7,-9,1).
    \]
    Applying the downward carry twice gives
    \[
        (7,-9,1)\longmapsto (6,-3,0)\longmapsto (5,3,1),
    \]
    so
    \[
        24\oplus_2 31=53.
    \]

    Likewise, for multiplication \autoref{prop:order2_multiplication_nosqrt} gives
    \[
        \lambda_2(12)=(2,1,0),\qquad \lambda_2(20)=(2,2,0).
    \]
    Hence the raw product triple is
    \[
        (M,U,E)=(4,2^2\cdot 2+2^2\cdot 1+2\cdot 1\cdot 2,0)=(4,16,0).
    \]
    Applying the upward carry rules twice gives
    \[
        (4,16,0)\longmapsto (5,12,1)\longmapsto (6,6,0),
    \]
    so
    \[
        12\otimes_2 20=60.
    \]
    In the first computation, the sum is found directly in the representation: the raw triple starts in layer $7$ and the carries move it left until it reaches the canonical odd representative $(5,3)$. In the second, the product starts in layer $4$ and the carries move it right until it reaches the terminal representative $(6,0)$. Thus, in order $2$, the onion representation supports direct addition and multiplication by layer normalization. A natural next step is to replace these layer-wise procedures by genuinely local rules, analogous to schoolbook arithmetic, and then to extend them to higher orders.
\end{example}

\subsection{Direct arithmetic in order \texorpdfstring{$3$}{3}}

The order-$3$ case has the same flavor, but each layer is traversed in three phases rather than by a zig-zag. The next proposition gives the exact layer order for the reverse prefer-max sequence.

\begin{proposition}[Explicit layer order for the reverse prefer-max sequence of order $3$]\label{prop:order3_explicit}
    Let $(x_i)_{i=0}^{\infty}$ be the reverse prefer-max onion sequence of order $3$, and let $\rho_3$ be its rank map. Fix $m\geq 1$. Then the words of layer $m$ appear in the following order.

    For each pair $(v,w)$ with $0\leq v<m$ and $0\leq w<m$, taken in lexicographic order, one has the triple
    \[
        (w,v,m),\qquad (v,m,(w+1)\bmod m),\qquad
        \begin{cases}
            (m,w+1,v), & w<m-1,\\
            (m,0,v+1), & w=m-1.
        \end{cases}
    \]
    After these $m^2$ triples, for each $w=0,\dots,m-2$ one has
    \[
        (w,m,m),\qquad (m,m,w+1),\qquad (m,w+1,m),
    \]
    and the layer ends with
    \[
        (m-1,m,m),\qquad (m,m,m),\qquad (m,m,0),\qquad (m,0,0).
    \]
\end{proposition}

\begin{proof}
    By the Fredricksen--Maiorana description of reverse prefer-max (see Theorem~6 of~\cite{amram2021cycle}), the order-$3$ sequence is obtained by concatenating, in colex order, the expansions $\ell^{3/|\ell|}$ of the Lyndon words $\ell$ whose lengths divide $3$. Thus a layer consists of the length-$3$ Lyndon words whose maximal symbol is $m$, followed by the length-$1$ Lyndon word $m$.

    We first characterize the length-$3$ Lyndon words in layer $m$. Let $xyz$ be such a word. Since some rotation of $xyz$ ends in $m$, while a colex-maximal word must itself be the colex-maximal rotation, necessarily $z=m$. If also $x=m$, then the rotation $ymx$ ends in $m$ and has middle letter $m$, so $ymx>_{colex}xym$, a contradiction. Hence $x<m$. Conversely, if $x<m$, $y\leq m$, and $z=m$, then $xym$ is larger than the rotation $ymx$ because their last letters are $m$ and $x$, and it is larger than the rotation $mxy$ because either $y<m$ and the last letters are $m$ and $y$, or else $y=m$ and the middle letters are $m$ and $x$. Therefore the length-$3$ Lyndon words in layer $m$ are exactly the words
    \[
        wvm\qquad (0\leq w<m,\ 0\leq v\leq m).
    \]

    Among these words, colex order is exactly the lexicographic order of the pair $(v,w)$: the last letter is always $m$, so the middle letter is compared first, and the first letter breaks ties. Thus the length-$3$ Lyndon words of layer $m$ are
    \[
        00m,10m,\dots,(m-1)0m,01m,\dots,(m-1)1m,\dots,0mm,\dots,(m-1)mm.
    \]

    Let $\ell=wvm$ and let $\ell'=w'v'm$ be the next word in this list. The three consecutive windows across the block $\ell\ell'$ are
    \[
        (w,v,m),\qquad (v,m,w'),\qquad (m,w',v').
    \]
    If $v<m$, then $(w',v')=(w+1,v)$ for $w<m-1$, while $(w',v')=(0,v+1)$ for $w=m-1$. This gives the first family of triples. If $v=m$ and $w<m-1$, then $\ell'=(w+1)mm$, which gives the second family.

    After the last length-$3$ Lyndon word $(m-1)mm$, the next Lyndon word is the length-$1$ word $m$, whose expansion is $mmm$. Hence the remaining windows are
    \[
        (m-1,m,m),\qquad (m,m,m),\qquad (m,m,0).
    \]
    Finally, the next layer begins with the Lyndon word $00(m+1)$, so the overlap with $mmm$ contributes the last word $(m,0,0)$.
\end{proof}

The explicit layer description now translates directly into closed rank and unrank formulas. The layer is determined by the maximal symbol $m$, the quotient-remainder decomposition $s=vm+w$ records the position of the underlying Lyndon block inside that layer, and the residue of the offset modulo $3$ specifies which word of the corresponding triple is being read. The next corollary packages this information into formulas for $\rho_3$ and $\rho_3^{-1}$.

\begin{corollary}[Explicit rank and unrank for order $3$]\label{cor:order3_rank_unrank}
    Let $\rho_3$ be the rank map of the reverse prefer-max onion sequence of order $3$. Denote by $S$ the set of $(a,b,c)$ that are of the form $m^{3-i}0^i$ for some $i\in\{0,1,2\}$ and some 
    $m\neq 0$. Then for every $a,b,c \in \mathbb{N}$ we have
    \[
        \rho_3(a,b,c)=
        \begin{cases}
            (m+1)^3-(3-i), & (a,b,c) \in S \\
            c^3+3bc+3a, & a<c \land b\leq c,\\
            a^3+3ac+3b-1, & b<a \land c\leq a,\\
            b^3+3ab+3((c-1)\bmod b)+1, & c<b \land a\leq b.
        \end{cases}
    \]
    Conversely, $\rho_3^{-1}(0)=(0,0,0)$. If $N\ge 1$, let
    \[
        m=\bigl\lfloor \sqrt[3]{N}\bigr\rfloor,\qquad
        t=N-m^3,\qquad
        \varepsilon=t\bmod 3,\qquad
        s=\Bigl\lfloor \frac t3 \Bigr\rfloor,
    \]
    \[
        w=s\bmod m,\qquad
        v=\Bigl\lfloor \frac sm \Bigr\rfloor,\qquad
        \delta=(w+1)\bmod m.
    \]
    Then
    \[
        \rho_3^{-1}(N)=
        \begin{cases}
            \text{the word } m^{3-i}0^i, & N=(m+1)^3-(3-i) \text{ for some } i\in\{0,1,2\},\\
            (w,v,m), & \varepsilon=0,\\
            (v,m,\delta), & \varepsilon=1,\\
            (m,0,v+1), & \varepsilon=2 \text{ and } \delta=0,\\
            (m,\delta,v), & \varepsilon=2 \text{ and } \delta\neq 0.
        \end{cases}
    \]
    Consequently, the operations
    \[
        (a,b,c)\oplus_3(d,e,f)=\rho_3^{-1}\bigl(\rho_3(a,b,c)+\rho_3(d,e,f)\bigr)
    \]
    and
    \[
        (a,b,c)\otimes_3(d,e,f)=\rho_3^{-1}\bigl(\rho_3(a,b,c)\rho_3(d,e,f)\bigr)
    \]
    are explicit addition and multiplication rules on the order-$3$ onion representations. Moreover, for $(d,e,f)\neq (0,0,0)$,
    \[
        Q_3((a,b,c),(d,e,f))=\rho_3^{-1}\left(\left\lfloor \frac{\rho_3(a,b,c)}{\rho_3(d,e,f)}\right\rfloor\right)
    \]
    and
    \[
        R_3((a,b,c),(d,e,f))=\rho_3^{-1}\bigl(\rho_3(a,b,c)\bmod \rho_3(d,e,f)\bigr)
    \]
    are explicit quotient and remainder rules on the order-$3$ onion representations.
\end{corollary}

\begin{proof}
    By \autoref{prop:order3_explicit}, the nonterminal words in layer $m$ occur at offsets
    \[
        3(vm+w),\qquad 3(vm+w)+1,\qquad 3(vm+w)+2,
    \]
    according to whether the word is of the form
    \[
        (w,v,m),\qquad (v,m,(w+1)\bmod m),\qquad (m,(w+1)\bmod m,v),
    \]
    with the third word interpreted as $(m,0,v+1)$ when $w=m-1$. This immediately gives
    \[
        \rho_3(w,v,m)=m^3+3(vm+w),
    \]
    which is the branch $c^3+3bc+3a$ after writing $(a,b,c)=(w,v,m)$.

    If $b<a$ and $c\leq a$, then $(b,c,a)$ is of the first type in layer $a$, so
    \[
        \rho_3(b,c,a)=a^3+3ac+3b.
    \]
    By \autoref{prop:order3_explicit}, the word $(a,b,c)$ occurs immediately before $(b,c,a)$, hence
    \[
        \rho_3(a,b,c)=a^3+3ac+3b-1.
    \]

    If $c<b$ and $a\leq b$, then the first word of the corresponding triple is
    \[
        \bigl((c-1)\bmod b,a,b\bigr),
    \]
    so
    \[
        \rho_3\bigl((c-1)\bmod b,a,b\bigr)=b^3+3ab+3((c-1)\bmod b).
    \]
    The word $(a,b,c)$ occurs immediately after it, which yields the branch
    \[
        \rho_3(a,b,c)=b^3+3ab+3((c-1)\bmod b)+1.
    \]

    The last three words of the layer are $(m,m,m)$, $(m,m,0)$, and $(m,0,0)$, so their indices are $(m+1)^3-3$, $(m+1)^3-2$, and $(m+1)^3-1$, respectively. This gives the terminal branch.

    Conversely, let $N\geq 1$ and write
    \[
        N=m^3+t,\qquad t=3s+\varepsilon,\qquad s=vm+w,
    \]
    with $\varepsilon\in\{0,1,2\}$ and $0\leq w<m$. The terminal cases are exactly the three values $(m+1)^3-3$, $(m+1)^3-2$, and $(m+1)^3-1$. Outside those cases, \autoref{prop:order3_explicit} shows that $\varepsilon=0$ gives the first word of the triple, namely $(w,v,m)$, $\varepsilon=1$ gives the second word, namely $(v,m,(w+1)\bmod m)$, and $\varepsilon=2$ gives the third word, namely $(m,(w+1)\bmod m,v)$, with the wrap case $(m,0,v+1)$ when $w=m-1$. This is exactly the displayed formula for $\rho_3^{-1}$. The formulas for $\oplus_3$, $\otimes_3$, $Q_3$, and $R_3$ are then immediate by transport of ordinary arithmetic through $\rho_3$.
\end{proof}

\autoref{cor:order3_rank_unrank} still describes arithmetic through the global rank map. As in order $2$, the next step is to work with canonical layer-offset coordinates and normalize addition and multiplication by finitely many carries.

\begin{definition}[Canonical order-$3$ coordinates]
    For a word $x$ in the reverse prefer-max onion sequence of order $3$, define $\lambda_3(x)=(m,u,\varepsilon)$ by
    \[
        \lambda_3(abc)=
        \begin{cases}
            (a+1,0,-1), & b=c=0, \\
            (a,a^2+a,-1), & a=b\land c=0, \\
            (a,a^2+a-1,1), & a=b=c, \\
            (c,bc+a,0), & c\geq b\land c>a, \\
            (b,ab+(c-1)\bmod b,1), & b\geq a\land b>c, \\
            (a,ac+b,-1), & a\geq c\land a>b.
        \end{cases}
    \]
    and notice that:
    \[
        \rho_3(x)=m^3+3u+\varepsilon,
    \]
    where $\varepsilon\in\{-1,0,1\}$ and the admissible range of $u$ is
    \[
        \begin{cases}
            0\leq u\leq m^2+m-1, & \varepsilon=1,\\
            0\leq u\leq m^2+m, & \varepsilon=0 \text{ or } \varepsilon=-1.
        \end{cases}
    \]
    Conversely, let $\lambda_3^{-1}(m,u,\varepsilon)$ denote the canonical word associated with such a triple, be defined like this:
    \[
        \lambda_3^{-1}(m,u,0)=(u\bmod m,\Big\lfloor\frac{u}{m}\Big\rfloor,m)
    \]
    \[
        \lambda_3^{-1}(m,u,1)=
        \begin{cases}
            (\lfloor\frac{u}{m}\rfloor,m,(u+1)\bmod m), & 0\leq u<m^2+m-1,\\
            (m,m,m), & u=m^2+m-1,
        \end{cases}
    \]
    and
    \[
        \lambda_3^{-1}(m,u,-1)=
        \begin{cases}
            (m-1,0,0), & u=0,\\
            (m,m,0), & u=m^2+m,\\
            (m,u\bmod m,\lfloor\frac{u}{m}\rfloor), & otherwise.
        \end{cases}
    \]
\end{definition}

Thus, if $\lambda_3(x)=(m,u,\varepsilon)$, then $m=\lfloor \sqrt[3]{\rho_3(x)}\rfloor$: the first coordinate is exactly the represented layer, while $u$ and $\varepsilon$ record the offset inside that layer.

\begin{proposition}[Direct addition for order $3$]\label{prop:order3_addition_nocuberoot}
    Let $\lambda_3(x)=(m,u,\varepsilon)$ and $\lambda_3(y)=(n,v,\eta)$. Define the raw sum triple by
    \[
        M=m+n,\qquad U=u+v-mn\cdot M+\Bigl\lfloor \frac{\varepsilon+\eta+1}{3}\Bigr\rfloor,\qquad E=((\varepsilon+\eta+1)\bmod 3)-1.
    \]
    Then
    \[
        \rho_3(x)+\rho_3(y)=M^3+3U+E.
    \]
    Moreover the raw triple cannot overshoot layer $M$, so it can fail to be canonical only on the low side. In that case one repeatedly applies the downward carry rule:
    \[
        (m,u,\varepsilon)\longmapsto (m-1,u+m^2-m+\Big\lfloor\frac{\varepsilon+2}{3}\Big\rfloor,(\varepsilon-1)\bmod3-1)\qquad \text{while } u<0,
    \]
    This process terminates at a unique canonical triple $(m,u,\varepsilon)$, and the corresponding sum word is
    \[
        x\oplus_3 y=\lambda_3^{-1}(m,u,\varepsilon).
    \]
    Hence order-$3$ addition can be carried out directly in the onion representation, without global rank inversion.
\end{proposition}

\begin{proof}
    Let
    \[
        N=\rho_3(x)+\rho_3(y).
    \]
    From $\rho_3(x)=m^3+3u+\varepsilon$ and $\rho_3(y)=n^3+3v+\eta$ we get
    \[
        N=m^3+n^3+3u+3v+\varepsilon+\eta.
    \]
    Since $m^3+n^3=(m+n)^3-3mn(m+n)$, and since
    \[
        \varepsilon+\eta
        =
        3\Bigl\lfloor \frac{\varepsilon+\eta+1}{3}\Bigr\rfloor
        +
        \bigl((\varepsilon+\eta+1)\bmod 3-1\bigr),
    \]
    this becomes
    \[
        N=M^3+3U+E,
    \]
    with $M,U,E$ as above.

    To see that there is no upper overflow, note that
    \[
        \rho_3(x)\leq (m+1)^3-1,\qquad \rho_3(y)\leq (n+1)^3-1,
    \]
    so
    \[
        N\leq (m+1)^3+(n+1)^3-2<(m+n+1)^3=(M+1)^3.
    \]
    Since $N=M^3+3U+E$ with $E\in\{-1,0,1\}$, this implies
    \[
        U\leq
        \begin{cases}
            M^2+M-1, & E=1,\\
            M^2+M, & E=0 \text{ or } E=-1.
        \end{cases}
    \]
    These are exactly the upper bounds for canonical triples in layer $M$, so the raw triple can fail to be canonical only on the low side, namely when $U<0$.

    The carry rules are the identities
    \[
        m^3+3u-1=(m-1)^3+3(u+m^2-m),
    \]
    \[
        m^3+3u=(m-1)^3+3(u+m^2-m)+1,
    \]
    and
    \[
        m^3+3u+1=(m-1)^3+3(u+m^2-m+1)-1.
    \]
    These are exactly the three cases $\varepsilon=-1,0,1$ of
    \[
        (m,u,\varepsilon)\longmapsto
        (m-1,u+m^2-m+\Big\lfloor\frac{\varepsilon+2}{3}\Big\rfloor,(\varepsilon-1)\bmod3-1).
    \]
    Hence each downward carry preserves the represented integer. If a current triple $(m,u,\varepsilon)$ has $u<0$, then
    \[
        N=m^3+3u+\varepsilon<m^3,
    \]
    so its canonical representative must lie in a smaller layer. Each carry lowers the layer by $1$, so the process must eventually stop.

    Let $(m,u,\varepsilon)$ be the first triple in the carry sequence with $u\geq 0$. Then
    \[
        N=m^3+3u+\varepsilon\geq m^3.
    \]
    If no carry was applied, then $m=M$ and the bound $N<(M+1)^3$ already gives $N<(m+1)^3$. Otherwise this triple was obtained from a previous triple $(m+1,u',\varepsilon')$ with $u'<0$, and then
    \[
        N=(m+1)^3+3u'+\varepsilon'<(m+1)^3.
    \]
    In either case,
    \[
        m^3\leq N<(m+1)^3,
    \]
    so $m=\lfloor \sqrt[3]{N}\rfloor$ is the true layer of the sum. Since $N=m^3+3u+\varepsilon$ and $\varepsilon\in\{-1,0,1\}$, the upper bound $N<(m+1)^3$ yields
    \[
        0\leq u\leq
        \begin{cases}
            m^2+m-1, & \varepsilon=1,\\
            m^2+m, & \varepsilon=0 \text{ or } \varepsilon=-1.
        \end{cases}
    \]
    Thus the stopping triple is canonical. Its uniqueness follows from the uniqueness of $\rho_3^{-1}$, and the sum word is therefore $\lambda_3^{-1}(m,u,\varepsilon)$.
\end{proof}

\begin{proposition}[Direct multiplication for order $3$]\label{prop:order3_multiplication_nocuberoot}
    Let $\lambda_3(x)=(m,u,\varepsilon)$ and $\lambda_3(y)=(n,v,\eta)$. Define
    \[
        M=mn,\qquad S=m^3\eta+n^3\varepsilon+\varepsilon\eta,
    \]
    \[
        U=m^3v+n^3u+3uv+u\eta+v\varepsilon+\Bigl\lfloor \frac {S+1}3\Bigr\rfloor,\qquad E=(S+1)\bmod 3-1.
    \]
    Then
    \[
        \rho_3(x)\rho_3(y)=M^3+3U+E.
    \]
    Since $\rho_3(x)\rho_3(y)\geq m^3n^3=M^3$, the raw triple cannot lie below layer $M$. Thus it can fail to be canonical only on the high side, in which case one repeatedly applies the upward carry rule:
    \[
        (m,u,\varepsilon)\longmapsto (m+1,u-m^2-m-\Big\lfloor\frac{2-\varepsilon}3\Big\rfloor,\varepsilon\bmod3-1)\qquad \text{while } u>m^2+m-\Big\lfloor\frac{\varepsilon+2}{3}\Big\rfloor,
    \]
    This process terminates at a unique canonical triple $(m,u,\varepsilon)$, and the corresponding product word is
    \[
        x\otimes_3 y=\lambda_3^{-1}(m,u,\varepsilon).
    \]
    Hence order-$3$ multiplication can also be carried out directly in the onion representation, without global rank inversion.
\end{proposition}

\begin{proof}
    Let
    \[
        N=\rho_3(x)\rho_3(y).
    \]
    Expanding the product gives
    \[
        N=(m^3+3u+\varepsilon)(n^3+3v+\eta)
    \]
    \[
        =m^3n^3+3(m^3v+n^3u+3uv+u\eta+v\varepsilon)+S.
    \]
    Since
    \[
        S
        =
        3\Bigl\lfloor \frac{S+1}{3}\Bigr\rfloor
        +
        \bigl((S+1)\bmod 3-1\bigr),
    \]
    this is exactly
    \[
        N=M^3+3U+E.
    \]
    Also,
    \[
        N\geq m^3n^3=M^3,
    \]
    so $U\geq 0$, and the raw triple cannot fail to be canonical on the low side.

    The three upward carry rules are the identities
    \[
        m^3+3u-1=(m+1)^3+3(u-m^2-m-1)+1,
    \]
    \[
        m^3+3u=(m+1)^3+3(u-m^2-m)-1,
    \]
    and
    \[
        m^3+3u+1=(m+1)^3+3(u-m^2-m).
    \]
    Hence each upward carry preserves the represented integer. Moreover, whenever a carry is applied, the new offset remains nonnegative:
    \[
        u-m^2-m-1\geq 0
        \quad\text{if }u>m^2+m,
    \]
    and
    \[
        u-m^2-m\geq 0
        \quad\text{if }u>m^2+m-1.
    \]
    Thus every triple in the carry sequence has the form $N=m^3+3u+\varepsilon$ with $u\geq 0$, and therefore lies on or above layer $m$.

    If one of the carry conditions holds, then the represented integer already lies in a larger layer. Indeed, in the three cases we have
    \[
        m^3+3u-1\geq m^3+3(m^2+m+1)-1=(m+1)^3+1,
    \]
    \[
        m^3+3u\geq m^3+3(m^2+m+1)=(m+1)^3+2,
    \]
    and
    \[
        m^3+3u+1\geq m^3+3(m^2+m)+1=(m+1)^3.
    \]
    So whenever a carry is applied, necessarily $N\geq (m+1)^3$, and the canonical representative must lie in a layer strictly larger than $m$. Each carry raises the layer by $1$, so the process must terminate after finitely many steps.

    Let $(m,u,\varepsilon)$ be the first triple for which no upward carry applies. Since $u\geq 0$, we have
    \[
        N=m^3+3u+\varepsilon\geq m^3.
    \]
    We claim that $N<(m+1)^3$. Otherwise $N\geq (m+1)^3$, and then:
    \[
        3u-1\geq 3m^2+3m+1
    \]
    if $\varepsilon=-1$, so $u\geq m^2+m+1>m^2+m$;
    \[
        3u\geq 3m^2+3m+1
    \]
    if $\varepsilon=0$, so $u\geq m^2+m+1>m^2+m$; and
    \[
        3u+1\geq 3m^2+3m+1
    \]
    if $\varepsilon=1$, so $u\geq m^2+m>m^2+m-1$.
    In every case one of the carry conditions would still hold, contradicting the choice of $(m,u,\varepsilon)$.

    Therefore
    \[
        m^3\leq N<(m+1)^3,
    \]
    so $m=\lfloor \sqrt[3]{N}\rfloor$ is the true layer of the product. Since no upward carry applies, we also have
    \[
        0\leq u\leq
        \begin{cases}
            m^2+m-1, & \varepsilon=1,\\
            m^2+m, & \varepsilon=0 \text{ or } \varepsilon=-1.
        \end{cases}
    \]
    Thus the stopping triple is canonical. Its uniqueness follows from the uniqueness of $\rho_3^{-1}$, and the product word is therefore $\lambda_3^{-1}(m,u,\varepsilon)$.
\end{proof}

\begin{proposition}[Carry complexity in orders $2$ and $3$]\label{prop:carry_complexity}
    For $d\in\{2,3\}$, let $\rho_d$ and $\lambda_d$ denote the corresponding rank and coordinate maps, and write
    \[
        \lambda_d(x)=(m,u,\varepsilon),\qquad \lambda_d(y)=(n,v,\eta).
    \]
    Let
    \[
        L_d(N)=\Bigl\lfloor \sqrt[d]{N}\Bigr\rfloor.
    \]
    Then the normalization procedure for addition uses exactly
    \[
        c_d^+(x,y)=m+n-L_d\bigl(\rho_d(x)+\rho_d(y)\bigr)
    \]
    carry updates, while the normalization procedure for multiplication uses exactly
    \[
        c_d^\times(x,y)=L_d\bigl(\rho_d(x)\rho_d(y)\bigr)-mn
    \]
    carry updates.

    In particular,
    \[
        0\leq c_d^+(x,y)\leq m+n,\qquad 0\leq c_d^\times(x,y)\leq m+n.
    \]
    Hence, in both orders, direct addition and multiplication require at most $m+n$ carry updates, and therefore only linearly many elementary arithmetic operations in the input layers.
\end{proposition}

\begin{proof}
    In \autoref{prop:order2_addition_nosqrt} and \autoref{prop:order3_addition_nocuberoot}, the raw sum triple starts in layer $m+n$. Each downward carry preserves the represented integer and lowers the layer by exactly $1$. The normalization stops precisely when the triple reaches the canonical layer of $\rho_d(x)+\rho_d(y)$, namely $L_d(\rho_d(x)+\rho_d(y))$. Therefore the number of addition carries is exactly
    \[
        m+n-L_d\bigl(\rho_d(x)+\rho_d(y)\bigr).
    \]

    Similarly, in \autoref{prop:order2_multiplication_nosqrt} and \autoref{prop:order3_multiplication_nocuberoot}, the raw product triple starts in layer $mn$. Each upward carry preserves the represented integer and raises the layer by exactly $1$. The normalization stops when the layer reaches the canonical layer of $\rho_d(x)\rho_d(y)$, namely $L_d(\rho_d(x)\rho_d(y))$. Hence the number of multiplication carries is exactly
    \[
        L_d\bigl(\rho_d(x)\rho_d(y)\bigr)-mn.
    \]

    The addition bound is immediate. For multiplication, since
    \[
        \rho_d(x)<(m+1)^d,\qquad \rho_d(y)<(n+1)^d,
    \]
    we have
    \[
        \rho_d(x)\rho_d(y)<((m+1)(n+1))^d.
    \]
    Therefore
    \[
        L_d\bigl(\rho_d(x)\rho_d(y)\bigr)<(m+1)(n+1)=mn+m+n+1,
    \]
    so
    \[
        c_d^\times(x,y)\leq m+n.
    \]
    This proves the claimed bounds.
\end{proof}

Quotient and remainder are explicit in orders $2$ and $3$ by transporting ordinary Euclidean division through rank and unrank. The next proposition records this construction together with the resulting layer bounds.

\begin{proposition}[Transported division and layer bounds in orders $2$ and $3$]\label{prop:division_layers}
    For $d\in\{2,3\}$, let $x$ and $y$ be words in the reverse prefer-max onion sequence of order $d$, with $\rho_d(y)>0$. Define
    \[
        q_d(x,y)=\left\lfloor \frac{\rho_d(x)}{\rho_d(y)}\right\rfloor,\qquad
        r_d(x,y)=\rho_d(x)\bmod \rho_d(y),
    \]
    and let
    \[
        Q_d(x,y)=\rho_d^{-1}\bigl(q_d(x,y)\bigr),
    \]
    \[
        R_d(x,y)=\rho_d^{-1}\bigl(r_d(x,y)\bigr).
    \]
    Write
    \[
        \lambda_d(x)=(m,u,\varepsilon),\qquad
        \lambda_d(y)=(n,v,\eta),
    \]
    \[
        \lambda_d(Q_d(x,y))=(s,p,\xi),\qquad
        \lambda_d(R_d(x,y))=(t,q,\zeta).
    \]
    Then
    \[
        \left\lfloor \frac{m}{n+1}\right\rfloor \leq s \leq \left\lfloor \frac{m}{n}\right\rfloor
    \]
    and
    \[
        0\leq t\leq n.
    \]

    In particular, quotient and remainder are explicit in orders $2$ and $3$ through rank transport. A direct carry-normalized division rule remains open.
\end{proposition}

\begin{proof}
    Let
    \[
        A=\rho_d(x),\qquad B=\rho_d(y),\qquad q=q_d(x,y),\qquad r=r_d(x,y).
    \]
    Since $\lambda_d(x)$ and $\lambda_d(y)$ have layers $m$ and $n$, \autoref{prop:rank_layer} gives
    \[
        m^d\leq A<(m+1)^d,\qquad n^d\leq B<(n+1)^d.
    \]

    For the upper bound on the quotient layer, we have
    \[
        q=\left\lfloor \frac AB\right\rfloor<\frac{(m+1)^d}{n^d}=\left(\frac{m+1}{n}\right)^d.
    \]
    Since $\frac{m+1}{n}\leq \lfloor m/n\rfloor+1$, it follows that
    \[
        q<\bigl(\lfloor m/n\rfloor+1\bigr)^d.
    \]
    Therefore \autoref{cor:layer_root} implies
    \[
        s=\left\lfloor \sqrt[d]{q}\right\rfloor\leq \left\lfloor \frac{m}{n}\right\rfloor.
    \]

    For the lower bound, let $\ell=\lfloor m/(n+1)\rfloor$. Then $\ell\leq m/(n+1)$, hence
    \[
        \ell^d\leq \frac{m^d}{(n+1)^d}<\frac AB.
    \]
    Since $\ell^d$ is an integer, this gives $q\geq \ell^d$, and therefore
    \[
        s=\left\lfloor \sqrt[d]{q}\right\rfloor\geq \ell=\left\lfloor \frac{m}{n+1}\right\rfloor.
    \]

    Finally, $0\leq r<B<(n+1)^d$, so another application of \autoref{cor:layer_root} gives
    \[
        t=\left\lfloor \sqrt[d]{r}\right\rfloor\leq n.
    \]
    This proves the claim.
\end{proof}

\begin{example}\label{ex:order3_arithmetic}
    \textup{For layer $m=2$, the formulas give the order}
    \[
        \begin{aligned}
            &002,\;021,\;210,\;102,\;020,\;201,\;012,\;121,\;211,\;112,\\
            &120,\;202,\;022,\;221,\;212,\;122,\;222,\;220,\;200.
        \end{aligned}
    \]
    \textup{For instance,}
    \[
        \rho_3(1,2,0)=2^3+3\cdot 1\cdot 2+3((0-1)\bmod 2)+1=18,
    \]
    \textup{while}
    \[
        \rho_3^{-1}(23)=(1,2,2)
    \]
    \textup{because $23=2^3+15$, so $\varepsilon=0$, $s=5$, $v=2$, and $w=1$. For direct addition, the new proposition gives}
    \[
        \lambda_3(120)=(2,3,1),\qquad \lambda_3(021)=(2,0,1),
    \]
    \[
        (M,U,E)=\left(4,3+0-2\cdot2\cdot4+1,-1\right)=(4,-12,-1).
    \]
    \textup{A single downward carry sends this to $(3,0,0)$, hence}
    \[
        120\oplus_3 021=003.
    \]
    \textup{Thus order $3$ admits not only explicit rank and unrank formulas, but also a genuine carry-normalization calculus parallel to the order-$2$ case.}
\end{example}

\section{A bounded switching experiment}\label{section:implementation_switching}

To test whether the locality of onion counting can translate into lower switching activity, we carried out a bounded experiment in a simple register-level model. We compared two counters with the same number of states, namely $6561=9^4$: a standard binary counter modulo $6561$, stored in a $13$-bit register, and the reverse prefer-max onion counter of order $4$ truncated at maximal symbol $8$. In the onion case, the state trajectory was generated by repeated application of the successor rule $\operatorname{succ}$ starting from $0000$; in this bounded range, that orbit is exactly the reverse prefer-max order on $[9]^4$.

The binary baseline is the usual one: the state is a register $b\in\{0,\dots,6560\}$, encoded in binary, and one increment performs $b\leftarrow b+1 \pmod{6561}$. We measure the Hamming distance between the old and new $13$-bit encodings.

For the onion counter we do not physically shift the four-symbol window. Instead, we store the current state as a circular buffer $(c_0,c_1,c_2,c_3)\in \{0,\dots,8\}^4$ together with a head pointer $h\in\{0,1,2,3\}$. Each symbol is encoded in $4$ binary wires, so the buffer uses $16$ wires in total, and the pointer uses $2$ wires. The logical onion word represented by this physical state is
\[
    (c_h,c_{h+1 \bmod 4},c_{h+2 \bmod 4},c_{h+3 \bmod 4}).
\]
If this logical word is $w$ and its onion successor is $w'$, then necessarily $w'[1..3]=w[2..4]$, so $w'$ is obtained by discarding the first symbol of $w$ and appending one new symbol. In the moving-pointer realization, this is implemented by writing that new symbol into the current head cell and then advancing the head, namely $c_h\leftarrow w'_4$ and $h\leftarrow h+1 \pmod 4$. Thus each increment touches exactly one symbol cell physically, even though the logical window advances by one position.

We considered two encodings of the head pointer. In the first, $h$ is stored in ordinary binary. In the second, it is stored in the $2$-bit Gray order $0\mapsto 00$, $1\mapsto 01$, $2\mapsto 11$, $3\mapsto 10$, so that every pointer increment changes exactly one pointer bit. This Gray coding is relevant only to the measured switching activity of the stored implementation state; it does not alter the logical onion order or the symbol-update rule.

We measured two quantities. First, we counted how many stored symbol cells actually change value on one increment. By construction, the moving-pointer onion counter touches at most one symbol cell per step. In the present bounded experiment, the overwritten cell keeps the same value in ${4570}/{6561}\approx 69.7\%$ of all increments, so an actual symbol-value change occurs in only ${1991}/{6561}\approx 30.3\%$ of the steps. By contrast, the binary and Gray-encoded rank counters change at least one stored bit on every increment, and on rare wraparound-type events they can change many bits at once.

Second, we measured the Hamming distance between successive encoded states. Since Gray coding is the standard low-Hamming baseline, we also evaluated the reflected Gray encoding of the rank counter, namely $g(b)=b\oplus (b \gg 1)$ on the same modulo-$6561$ orbit. The full-state comparisons are summarized in \autoref{tab:switching_stats}, while the data-field locality comparison is collected separately in \autoref{tab:data_locality_stats}.

\begin{table}[ht]
    \centering
    \small
    \begin{tabular}{lccc}
        \hline
        Measured encoding & Average toggles & Worst case & Peak/average \\
        \hline
        Binary counter state & 1.9997 & 13 & 6.50 \\
        Gray-encoded rank counter & \textbf{1.0008} & 6 & 6.00 \\
        Onion full state, binary pointer & 2.0072 & 6 & \textbf{2.99} \\
        Onion full state, Gray pointer & 1.5072 & \textbf{5} & 3.32 \\
        \hline
    \end{tabular}
    \caption{Full-state switching statistics in the bounded experiment. Smaller is better in every numeric column.}
    \label{tab:switching_stats}
\end{table}

\begin{table}[ht]
    \centering
    \small
    \begin{tabular}{lcccc}
        \hline
        Field & Local? & Avg. & Worst & Change rate \\
        \hline
        Binary word & no & 1.9997 & 13 & 100\% \\
        Gray rank word & no & 1.0008 & 6 & 100\% \\
        Onion symbol field & \textbf{yes} & \textbf{0.5072} & \textbf{4} & \textbf{30.3\%} \\
        \hline
    \end{tabular}
    \caption{Data-field locality and switching statistics. The onion row excludes the head pointer.}
    \label{tab:data_locality_stats}
\end{table}

These numbers separate two design goals. If one only wants to minimize the average Hamming distance of a flat full-state encoding, then the Gray-encoded rank counter is best in \autoref{tab:switching_stats}. The onion implementation targets a different operating point: locality of writes together with bounded switching bursts. On the full stored state, a Gray-coded head pointer keeps the worst case at $5$, below both the Gray rank counter's $6$ and the binary counter's $13$, while the binary-pointer onion state achieves the smallest peak-to-average ratio, $2.99$. On the wide data path itself, \autoref{tab:data_locality_stats} shows a stronger form of superiority: the onion symbol field is the only representation with a local successor rule, it has the smallest average and worst-case switching, and it changes at all in only ${1991}/{6561}\approx 30.3\%$ of the steps. Thus, the experiment does not claim average-toggle optimality in the Gray-code sense; rather, it isolates a different advantage: strictly local state updates with a better worst-case switching bound and a calmer wide-state data path. This may also be relevant in side-channel-sensitive settings, including cryptographic counters and schedulers, where unusually large full-register updates can have a distinctive power-consumption signature and may therefore leak information through switching activity. The experiment is reproducible by the script \texttt{switching\_activity.py} described in the appendix.

\section{Conclusions and future directions}

Onion De Bruijn sequences have both a rigid side and a flexible side. Structurally, every onion sequence decomposes into layers indexed by the maximal symbol, and the onion condition forces the boundaries between those layers. Enumeratively, \autoref{thm:layer_enumeration} and \autoref{cor:onion_prefix_count} show that each layer admits many Hamiltonian orders and that these choices are independent once the boundary words are fixed. Reverse prefer-max is therefore distinguished not by uniqueness, but by the extra arithmetic regularity of one compatible family of layer orders.

That regularity already yields arithmetic at the structural level. \autoref{prop:rank_layer}, \autoref{cor:layer_root}, \autoref{cor:layer_carries}, and \autoref{cor:dominant_digit} show that the maximal symbol determines the exact layer of the represented integer, that $\lfloor \sqrt[n]{N}\rfloor$ can be read off directly from that layer, that perfect $n$th powers are the carry thresholds between layers, and that dominant symbols are monotone in fixed contexts. For the reverse prefer-max sequences of orders $2$ and $3$, \autoref{prop:order2_explicit}, \autoref{cor:order2_rank_unrank}, \autoref{prop:order2_addition_nosqrt}, \autoref{prop:order2_multiplication_nosqrt}, \autoref{prop:order3_explicit}, \autoref{cor:order3_rank_unrank}, \autoref{prop:order3_addition_nocuberoot}, and \autoref{prop:order3_multiplication_nocuberoot} sharpen this into explicit layer orders, rank and unrank formulas, and direct carry-normalized arithmetic. \autoref{prop:carry_complexity} then makes the algorithmic content explicit by giving exact carry counts and linear carry complexity in both orders, while \autoref{prop:division_layers} shows that quotient and remainder are also explicit by rank transport and satisfy natural layer bounds.

This arithmetic differs from more classical non-standard numeration systems. In mixed-radix systems, including the factorial number system, carries move between predetermined positions with prescribed radices~\cite{ercegovac2004digital}. In word-based systems such as abstract numeration systems and Ostrowski numeration~\cite{lecomte2010abstract,hieronymi2018ostrowski}, integers are ordered through a language and arithmetic is often studied automata-theoretically. Onion arithmetic combines aspects of both settings, but with a crucial extra constraint: the representing words must appear as consecutive windows of a single infinite De Bruijn sequence. The main carry mechanism therefore moves between layers of a fixed-length window, rather than between positions.

Onion numeration is not a universal replacement for decimal or binary arithmetic; rather, it is a fixed-window counting model tailored to local successor updates and coarse magnitude queries. The state can remain in an $n$-cell shift register throughout the computation, while \autoref{cor:layer_root} gives $\lfloor \sqrt[n]{N}\rfloor$ by inspection, perfect $n$th powers sit at layer boundaries, \autoref{prop:division_layers} gives explicit quotient and remainder with controlled layers, and in orders $2$ and $3$ the canonical coordinates support direct carry-normalized addition and multiplication. The bounded switching experiment of \autoref{section:implementation_switching} shows that this locality also has a concrete implementation consequence: the onion symbol field changes in only ${1991}/{6561}\approx 30.3\%$ of the steps, while the full stored state has worst-case switching $5$, compared with $6$ for the Gray-encoded rank counter and $13$ for the binary counter.

Taken together with the switching experiment in \autoref{section:implementation_switching}, these observations suggest several concrete directions for further study:
\begin{itemize}
    \item Shift-register counters and program counters, where the state remains a fixed-length word and the successor is implemented by local combinational logic;
    \item Low-switching address generators and sequencers, where adjacent states overlap heavily and layer changes mark coarse magnitude thresholds;
    \item Side-channel-aware cryptographic counters and state machines, where a more uniform update profile may help reduce leakage from transitions that would otherwise trigger conspicuously large power spikes;
    \item Streaming controllers and schedulers that benefit from immediate access to the current layer, and therefore to $\lfloor \sqrt[n]{N}\rfloor$;
    \item Analog or quantum implementations with adaptive-resolution state preparation and measurement, where small values are handled by coarse binary writing and discrimination, while larger layers trigger finer ternary or $n$-ary state setting and sampling that may consume additional energy or other resources only when that extra resolution is needed;
    \item Specialized arithmetic-on-state devices, especially in orders $2$ and $3$, where addition and multiplication can already be carried out directly in the onion representation.
\end{itemize}
We do not present these as finished applications; rather, they indicate where the combination of fixed-window counting, local shift rules, and layer-based arithmetic may merit further investigation.

The finite-alphabet literature already supplies much of the relevant toolkit. The Fredricksen--Maiorana theorem identifies the prefer-min De Bruijn sequence with the concatenation, in lexicographic order, of the Lyndon words whose lengths divide $n$~\cite{fredricksen1978necklaces}; see also~\cite{amram2018bruijn2} for a shift-rule proof. On the algorithmic side, \cite{amram2019shift},~\cite{amram2018bruijn2}, \cite{amram2020gsr}, and~\cite{sawada2017ranking} provide shift rules, jump rules, and ranking and unranking procedures for fixed alphabets. The next step is to make that toolkit compatible with growing alphabets. In particular, one would like successor and jump rules for the infinite reverse prefer-max order, and higher-order rank and unrank descriptions that expose the correct intra-layer offsets. Quotient and remainder are already explicit in orders $2$ and $3$ by transport through rank and unrank, but the central open problem is to understand those offsets well enough to obtain genuinely local addition, multiplication, and direct division on words of fixed length $n$.

\bibliographystyle{plain}
\bibliography{paper}

\newpage

\appendix

\section{Computational verification and software}

The code supplementing this paper is available in a companion repository at \url{https://github.com/geraw/onion-de-bruijn}. This appendix summarizes the contents of that repository and the scope of the checks that were carried out.

\paragraph{Available scripts.}
The file \texttt{order2\_debruijn\_arithmetic.py} implements the order-$2$ arithmetic developed in the paper: the rank map $\rho_2$, its inverse, the canonical coordinate map $\lambda_2$, the reconstruction map from canonical coordinates, the direct addition and multiplication procedures obtained by carry normalization, and quotient and remainder via rank transport. The code file \texttt{order3\_debruijn\_arithmetic.py} provides the analogous order-$3$ tools, including rank/unrank, canonical coordinates, direct addition and multiplication, and transported division.

The file \texttt{switching\_activity.py} evaluates the bounded implementation experiment discussed in \autoref{section:implementation_switching}. It compares a binary counter modulo $9^4$, the reflected Gray encoding of that rank counter, and the order-$4$ onion counter truncated at maximal symbol $8$, generates the onion orbit by iterating the current successor rule from $0000$, checks that this orbit agrees exactly with the reverse prefer-max order on $[9]^4$, uses a moving-pointer realization of the onion state, and reports the resulting symbol-write, bit-toggle, and peak-to-average burst statistics for both binary and Gray-coded head pointers.

\paragraph{Verification procedure.}
The \texttt{tests/} directory contains a multitude of tests for all order-2 and 3 arithmetic. Every file has 3 test suites: \texttt{Sanity}, \texttt{EdgeCases} and \texttt{All}. For order-2, the \texttt{Sanity} suite checks all examples from the paper, \texttt{EdgeCases} checks around layer boundaries and special words, and \texttt{All} checks all states up to $(2,36)$. For order-3, the \texttt{Sanity} and \texttt{EdgeCases} work similarly, and \texttt{All} checks all states up to $(2,16)$. \texttt{test\_hamiltonian\_paths.py} has individual test functions for the first section in the paper. To verify, check the \texttt{README.md} file in the repo (in short: use pytest and run the tests).

\paragraph{Use by readers.}
These scripts are meant to provide readers with working arithmetic models of the onion counting system. They can be used to
\begin{enumerate}
    \item convert between integers and their order-$2$ or order-$3$ onion representations,
    \item compute canonical layer-offset coordinates,
    \item perform direct addition and multiplication together with transported quotient and remainder inside the representation, and
    \item reproduce the finite computational checks by running the verification script.
\end{enumerate}
The verification code is included as a reproducibility aid for the explicit constructions, examples, and finite checks discussed in the paper.

\end{document}